

\input mtexsis
\paper
\tenpoint
\input epsf

\Eurostyletrue
\def\alfs{{\alpha_{0}}}
\def\alfp{{\alpha_{1}}}
\def\alfse{{\alpha_{2}}}
\def\lom{{\log({-s\over{\mu^2}})}}

\def\leb{{\biggl\lbrack}}
\def\les{{\biggl\lbrace}}
\def\rib{{\biggr\rbrack}}
\def\ris{{\biggl\rbrace}}
\def\cden{{16 \pi^2}}
\def\cof{{C_1^f}}
\def\ctf{{C_2^f}}
\def\coh{{C_1^H(\mu)}}
\def\cth{{C_2^H(\mu)}}
\def\argb{{{{M_H^2\over M_f^2}}}}
\def\arga{{a}}
\def\nc{{N_c}}
\def\arglog{{{\mu^2\over M_H^2}}}
\def\bi{{\beta_{i}}}
\def\gi{{B_{i}}}
\def\go{{B_{1}}}
\def\bs{{\beta_{0}}}
\def\bp{{\beta_{1}}}
\def\be{{\beta_{2}}}
\def\bsd{{ ({{-7}\over{18\pi}})(\alpha_{0})^2 }}
\def\bpd{{ ({1\over\pi})(\alpha_{1})^2        }}
\def\bed{{ ({{-11}\over{9\pi}})(\alpha_{2})^2 }}\referencelist
\reference{bl} J.L. Basdevant and B.W. Lee, \journal Phys. Rev. D;2,1680 
                 (1970)
\endreference
\reference{*bla} H. Lehmann, Acta Phys. Austriaca Suppl. XI (1973) 139
\endreference
\reference{dr} D.A. Dicus and W.W. Repko, \journal Phys. 
                 Rev. D;42,3660 (1990); \journal D;44,3473 (1992)
\endreference
\reference{*dra} D.A. Dicus and W.W. Repko, preprint CPP-5, To be published
\endreference
\reference{*drb}  D. Atkinson, M. Harada, and A.I. Sanda, 
                  \journal Phys. Rev. D;46,3884 (1992)
\endreference
\reference{wi}  R.S. Willey, \journal Phys. Rev. D;44,3646 (1991)
\endreference
\reference{*wia} H. Veltman and M. Veltman, 
              \journal Acta Phys. Pol. B;22,669 (1991) 
\endreference
\reference{tru}  T.N. Truong, \journal Phys. Rev. Lett.;70,888 (1993)
\endreference
\reference{chan} For a review, see M. Chanowitz, \journal Ann. Rev. Nuc. Sci;
                  38,323 (1988)
\endreference
\reference{hn} P. Hasenfratz and J. Nager, \journal Z. Phys. C;37,487 (1988)
\endreference
\reference{pt}  M.E. Peskin and T. Takeuchi, 
                \journal Phys. Rev. Lett.;65,964 (1990)
\endreference
\reference{fls}  A.F. Falk, M. Luke, and E.H. Simmons, 
                 \journal Nucl. Phys. B;365,523 (1991)
\endreference
\reference{gross} See, for example, D.J. Gross, preprint, PUPT-1329, 
                   hep-ph/9210207
\endreference
\reference{dv}  S. Dawson and G. Valencia, \journal Nucl. Phys. B;348,23 (1991)
\endreference
\reference{drg}  J.F. Donoghue, C. Ramirez, and G. Valencia, \journal Phys. 
               Rev. D;39,1947 (1988)
\endreference
\reference{*drga} J. Bagger, S. Dawson, and G. Valencia, fermilab preprint
                PUB-92/75-T, hep-ph/9204211\overfullrule=0pt 
\endreference
\reference{wein}  S. Weinberg, \journal Phys. Rev. Lett.;65,1177 (1990)
\endreference
\reference{*weina} T.N. Pham, \journal Phys. Lett. B;255,451 (1991)
\endreference
\reference{chan2} M. Chanowitz, preprint LBL-32938 (1992)
\endreference
\reference{bc}  S.R. Beane and C.B. Chiu, preprint CPP-93-08, hep-ph/9303254
\endreference
\reference{csb}  C.B. Chiu, E.C.G. Sudarshan, and G. Bhamathi, \journal Phys. 
                Rev. D;45,884 (1992)
\endreference
\reference{*csba} See also, S. Weinberg, \journal Nucl. Phys. B;363,3 (1991)
\endreference
\reference{ksrf}  K. Kawarabayashi and M. Suzuki, \journal Phys. 
                     Lett. B;16,225 (1966)
\endreference
\reference{*ksrfa} Riazuddin and Fayazuddin, \journal Phys. 
                      Rev. ;147,1071 (1966)
\endreference
\reference{mms}  See, for example, B.R. Martin, D. Martin, and G. Shaw, 
            Pion-Pion Interactions in Particle Physics (Academic,
London, 1976) and references therein
\endreference
\reference{col} S. Coleman, R. Jackiw, and H.D. Politzer, \journal Phys. Rev. D
		;10,2491 (1974)
\endreference
\reference{will}  S. Willenbrock, \journal Phys. Rev. D;43,1710 (1990)
\endreference
\endreferencelist
\titlepage
\title
Strongly Interacting W's and Z's and the Existence of a 
Heavy Fourth Generation of Fermions
\endtitle
\author
S.~R.~Beane\footnote\dag{{\rm sbeane@ccwf.cc.utexas.edu}} and S.~Varma\footnote\S{{\rm varmint@ccwf.cc.utexas.edu}}
Center for Particle Physics 
and Department of Physics  
The University of Texas at Austin
Austin, TX 78712
\endauthor
\abstract
\doublespaced

By employing the dictum that axiomatic principles are devoid of
predictive power, we find that the elastic unitarity constraint,
applied to strong W$_L$W$_L$ scattering, does not alter the assumed
spectrum of intermediate states.  We consider intermediate states
involving a heavy Higgs and heavy fermions of a hypothetical fourth
generation doublet. In contrast to recent studies, we find no p-wave
resonance, and therefore no violation of the S parameter upper bound.
We conclude that the elastic unitarity constraint sheds no light on
the existence of a heavy fourth generation.

\endabstract
\endtitlepage
\vfill\eject
\doublespaced
\superrefsfalse
Long ago, it was of interest to determine whether a strongly
interacting linear sigma model (LSM), in all its apparent simplicity,
could prove capable of generating the complex hadronic
spectrum\ref{bl}. In recent times, the familiar isomorphism between
the pions and the longitudinal modes of the standard model gauge
bosons, made precise through the equivalence theorem, has led to a
resurgence of interest in the strongly interacting
LSM\ref{dr}\ref{wi}\ref{tru}.  Motivation rests on the understanding
that W$_L$W$_L$ scattering provides a unique probe of the mechanism
responsible for electroweak symmetry breaking\ref{chan}.  Evidently,
triviality bounds\ref{hn} provide at best a narrow window within which
a strongly interacting Higgs sector could exist. However, it is important to
know what resonance structure to expect should such a window exist.
The specific question that we address is: Do intermediate states
involving heavy fermions of a hypothetical fourth generation doublet
provide enough binding to produce a p-wave resonance?  As emphasized
by Truong in \Ref{tru}, this question is of special interest since
precision weak-interaction measurements constrain the spin-1 content
of a strongly interacting Higgs sector via the S parameter upper
bound\ref{pt}.

Study of the singularity structure of the scattering amplitude
requires trading crossing symmetry for elastic unitarity, in a
non-unique way. In \Ref{wi} and \Ref{tru}, the method of Pad\'e
approximants is used to show that, for a large fermion mass, it is
possible to dynamically generate a p-wave resonance. If this result is
correct, then the S parameter bound can serve to exclude a heavy
fourth generation of fermions\ref{tru}. We will argue that the
use of the Pad\'e method in Refs.\refrange{bl}{tru} is based on the
notion that elastic unitarity should be imposed for the purpose of
making predictions.  Our approach is conceptually novel in that, in
sync with current lore, we ensure that unitarity per se yields no
predictive power, a point of view clearly orthogonal to S-matrix
theory (in the bootstrap sense.)  That axiomatic constraints like
unitarity and causality do not uniquely determine S-matrix elements
was an important lesson learned with the advent of QCD. A priori,
there are an infinite number of S-matrices consistent with the most
general physical principles\ref{gross}.  For example, in the context
of a non-abelian gauge field theory, changing gauge group and fermion
content certainly does not affect the unitarity of the theory.

We find that the I=1 singularity structure is insensitive to the heavy
fermion mass. Furthermore, the only nearby pole of the full amplitude
is seen to be the physical Higgs pole. Therefore, we find that there
is no violation of the S parameter upper bound for any value of the
heavy fermion mass.  We conclude that elastic unitarity, imposed as a
constraint on strong W$_L$W$_L$ scattering, yields no information
concerning the existence of a heavy fourth generation of fermions.

Exploitation of the model-independent low-energy structure of the theory is 
essential to our approach. Assuming a custodial SU(2) symmetry,
the most general effective Lagrangian including terms with four derivatives is
given by$^1$\vfootnote1{The coefficients, normalized in this way, are of
$O$(1) in the sense of naive dimensional analysis\ref{fls}.}

$$
\EQNalign{
{\cal L}&={{v^2} \over 4}{\Tr(\partial_\mu\Sigma\partial^\mu\Sigma^\dagger)}
	+{{C_1\over \cden}\Tr(\partial_\mu\Sigma\partial^\mu\Sigma^\dagger)
                       \Tr(\partial_\nu\Sigma\partial^\nu\Sigma^\dagger)}\cr
	&+{{C_2\over\cden}\Tr(\partial_\mu\Sigma\partial_\nu\Sigma^\dagger)
      \Tr(\partial^\mu\Sigma\partial^\nu\Sigma^\dagger)}.
                 \EQN chi\cr}
$$
The Goldstone boson fields ($w^+$,$w^-$, and $z$) are contained within the 
field variable ${\Sigma}={exp({{i \vec\tau\cdot\vec w}\over{v}})}$. 
$C_1$ and $C_2$ are undetermined constants which characterize the 
underlying theory at low energies. 
In general, there are contributions to $C_1$ and $C_2$ from all heavy
degrees of freedom, as well as continuum contributions arising from
goldstone boson loops.
The contributions to these low-energy
constants arising from intermediate states involving the Higgs boson and 
degenerate$^2$\vfootnote2{ The heavy fermions are taken to be degenerate
in order to avoid introducing isospin breaking terms.} heavy fermions of a 
fourth generation doublet have been calculated perturbatively in \Ref{dv}
using an on-shell subtraction scheme. 
They are given by

$$
\EQNalign{
\coh &={1\over 4} [-({9\pi\over 4 \sqrt3}-{37\over 9})-{1\over 6} \log(\arglog)]+2 \pi^2 ({v^2\over M_H^2})\cr
\cth &={1\over 4} [-({2\over 9})-{1\over 3} \log(\arglog)],
	\EQN piss\cr}
$$ and
$$
\EQNalign{
\cof &=-{\nc\over 12} [{1\over 2}+
	6 ({2\over\arga}+{(4-\arga)\over\arga^2} \int^1_0 \log(1-\arga x (1-x)) d\!x)]\cr
\ctf &={\nc\over 12},
	\EQN shit \cr} $$ where $\arga\equiv\argb$.  Note that for
definiteness we use values of the low-energy constants extracted from
perturbation theory. However, we stress that we could equally well
consider the most general couplings of fields with {\it any} quantum
numbers to the goldstone bosons, and estimate the values of these
couplings using naive dimensional analysis.  The uncertainty
associated with a change of the $C_{i}$ of $O$(1) should certainly not
exceed the inherent uncertainty that accompanies any unitarization
scheme. In fact, we find that our basic conclusions are insensitive to
natural changes in scale.  For example, we can replace the $C_{i}^H$
by the values that obtain from coupling a scalar to the goldstone
bosons in the most general way\ref{drg}.  In this case there is an
undetermined parameter that can be related to the scalar width. If,
instead of choosing the perturbative standard model value for the
width, we choose one-half of that value, as is the case when the
existence of a narrow p-wave resonance is assumed\ref{wein}, our
results are unaffected.

To order s$^2$, the relevant partial wave amplitudes of definite custodial 
isospin are given by

$$\EQNalign{
a_{0}(s)\equiv{a_{00}(s)}=&{\alpha_{0}s}\les 1-{{\alpha_{0}s}\over{\pi}}
                   \leb\log({-s\over{\mu^2}})-6(2C_1+C_2)\rib \cr
                         &-{{\alpha_{0}s}\over{\pi}}
                         \leb{7\over18}\log({s\over{\mu^2}})-{11\over108}
                         -{2\over3}(4C_1+5C_2)\rib\ris ,
\EQN chi5;a \cr
a_{1}(s)\equiv{a_{11}(s)}=&{\alpha_{1}s}\les 1-{{\alpha_{1}s}\over{\pi}}
                         \leb\log({-s\over{\mu^2}})-12C_2\rib \cr
                         &+{{\alpha_{1}s}\over{\pi}}
                         \leb\log({s\over{\mu^2}})-{1\over3}
                         +12(C_2-4C_1)\rib\ris , 
\EQN chi5;b \cr
a_{2}(s)\equiv{a_{20}(s)}=&{\alpha_{2}s}\les 1-{{\alpha_{2}s}\over{\pi}}
                         \leb\log({-s\over{\mu^2}})-12C_2\rib \cr
                         &-{{\alpha_{2}s}\over{\pi}}
                         \leb{11\over9}\log({s\over{\mu^2}})-{25\over54}
                         -{4\over 3}(8C_1+7C_2)\rib\ris,   
\EQN chi5;c \cr   }
$$
where
$\alfs\equiv{1\over{16\pi{v^2}}},\quad
  \alfp\equiv{1\over{96\pi{v^2}}},\quad{\rm and}\quad
  \alfse\equiv {-1\over{32\pi{v^2}}}$.
Each curly bracket consists of three terms, corresponding to the low-energy
theorem, and the $O$(s$^2$) contributions in the direct- and the 
crossed-channel respectively. {\it Note that we have been careful to preserve 
the crossing properties of the undetermined coefficients}\ref{bc}.

Our unitarization scheme corresponds to a simple bubble-sum with amplitude
given by

$$t_{i}(s)={{\alpha_i s}\over{1+{{\alpha_i s}\over\pi}[\lom+
     R_{i}(\mu^2)]}}. \EQN rh5 $$
The $R_{i}$'s are obtained by matching against the {\it direct-channel} piece 
of the chiral expansion. By inspection of \Eq{chi5} we find
$R_{0}={-6}(2C_1+C_2)$ and $R_{2}=R_{1}=-12C_2$. The ``complementarity''
between the I=1 and I=2 channels that follows from 
$R_{2}=R_{1}$ is investigated elsewhere in detail\ref{chan2}\ref{bc}.

Inspection of \Eq{shit} reveals that the I=1 and I=2 amplitudes are
independent of the heavy fermion mass, in sharp contrast with the
Pad\'e result of \Ref{wi} and \Ref{tru}.  Only the I=0 amplitude has
non-logarithmic contributions that depend on the Higgs and fermion
masses. This is not surprising; the values of $C_1$ and $C_2$ given in
\Eq{shit} are the low-energy manifestation of a
scalar-dominated theory. Unitarization simply restores the basic
properties of the assumed underlying theory.

In Fig.~1 we
schematically depict the complex s-plane. With a rather conservative
choice of cutoff, given by $\Lambda$=$4\pi v\simeq$ (3 TeV), and with
M$_H$=M$_f$= 1 TeV, we see that the only pole in the theory is the
``physical'' Higgs boson. In
Fig.~2 we display the partial wave amplitudes of definite
custodial isospin for values of the tree Higgs mass of 0.75 TeV and 1
TeV.  For values of M$_f$ above 250 GeV, the fermionic contributions
to the I=0 amplitude amount to a negligible renormalization of the
physical Higgs mass, and so we neglect them in the
graph. The complementary character of the non-resonant I=1 and I=2
amplitudes is clearly evident. We also display the Pad\'e prediction for
the I=1 amplitude, with M$_H$=0.75 TeV and M$_f$=1 TeV.

The approximation of neglecting crossed-channel contributions clearly
works best near an s-channel pole. Since our primary goal is to
investigate the possibility of a p-wave resonance for definite values
of C$_1$ and C$_2$, this sort of approximation is ideally suited to
the task.  More importantly, we argue that if one wants to play the
unitarization game, then one is {\it required} to make this
approximation. We have argued that no S-matrix element should be
uniquely determined by unitarity alone. Yet, we see in \Eq{rh5} that
if $t_{1}$ is resonant, the width of the resonance is automatically
fixed to the weak scale analogue of the KSRF relation\ref{wein}.
However, we need not worry.  This prediction is not a consequence of
imposing elastic unitarity, but rather of neglecting the left-hand
cut.  This is easily seen by including left-hand cut contributions in
a way that respects the low-energy structure of
\Eq{chi5}, and yet avoids double-counting of graphs\ref{csb}. 
\Eq{rh5} then becomes

$$t_{i}(s)={{\alpha_{i}s+{\bi}s^2[\log({s\over{\mu^2}})+{\gi}({\mu^2})]} \over{
1+{{\alpha_{i} s}\over\pi}[\lom+R_{i}(\mu^2)]+
{\bi\over{2\pi}}s^2[(\lom)^2+2\gi({\mu^2})\lom+ M_{i}({\mu^2})]}},
\overfullrule=0pt \EQN ep6$$ 
where $\bs\equiv\bsd$, $\bp\equiv\bpd$, and $\be\equiv\bed$ (see
\Eq{chi5}.)  The $\gi$ are the low-energy constants associated with heavy
particle exchange in the crossed-channel. The $M_{i}$ are undetermined
constants that appear at two-loop order in the chiral expansion.  We
see that it is by neglecting the contribution to the imaginary part of
the inverse amplitude involving $\go$ that we are able to predict the
KSRF relation.  Therefore, the predictive power of \Eq{rh5} is not a
result of imposing unitarity, but rather a result of neglecting a
class of graphs associated with heavy particle exchanges in the
crossed-channel, which are manifest at $O$(s$^2$) in the chiral
expansion. It is important to note that the above does not constitute
a new derivation of the KSRF relation. In fact, all justifications
of the KSRF relation, including the original current algebra
derivation\ref{ksrf}\relax, {\it require} the tacit assumption that
the left-hand cut of the I=1 scattering amplitude is
effectively absent\ref{mms}. We find it powerful evidence in favor of
our scheme that, by ensuring that predictive power come from a
source other than elastic unitarity, we arrive at a consistent
derivation of the KSRF relation.

The method of Pad\'e approximants, as applied in
Refs.\refrange{bl}{tru}, also predicts the KSRF relation in the I=1
channel, and yet the $O$(s$^2$) crossed-channel contributions are {\it
included}.  Therein lies its downfall; the neglect of crossed-channel
contributions can no longer serve as the source of predictive power,
and so the crossed-channel contributions necessarily appear in the
wrong place.  Yet if this is the case, then why do both unitarization
schemes of the bubble-sum type and the Pad\'e method provide a good
parametrization of the $\pi$-$\pi$ phase shift data? The reason is
straightforward.  One can say that the bubble-sum method works well
because the crossed-channel contributions which are neglected are
small, whereas the Pad\'e method works well because the
crossed-channel contributions {\it which are included in the wrong
place} are small.  Since these misplaced contributions appear in the
real part of the inverse amplitude, in the current context it is quite
understandable that unphysical poles are present.  Of course, there
are well defined instances in field theory where crossed-channel
contributions decouple. For example, the O(N) model is exactly
solvable to leading order in ${1\over N}$ precisely because left-hand
cut contributions first appear at $O({1\over N^2})$\ref{col}; not
surprisingly, to leading order in ${1\over N}$, the [1,1] Pad\'e
approximant yields the exact result\ref{will}. In this spirit, it is
interesting to note that if we assume that the crossed-channel
contributions that appear at $O(s^2)$ in \Eq{chi5} are much smaller
than the direct-channel contributions at the same order, then our
unitary amplitude, \Eq{rh5}, {\it is} the [1,1] Pad\'e approximant of
\Eq{chi5}.  However, our unitary amplitude with crossed-channel
contributions included, \Eq{ep6}, is clearly unrelated to any Pad\'e
approximant.  The moral of this story is that the Pad\'e method, which
is ideally suited to problems in potential theory, should be applied
only with great care to problems where crossing symmetry is important.

Our conclusions are not surprising.  The effective field theory
viewpoint implies that one gets out essentially what one puts in. Once
we saturate the low-energy constants of chiral perturbation theory
with contributions from a scalar-dominated underlying theory,
information regarding the intermediate-energy spectrum is, in a sense,
exhausted.  The elastic unitarity constraint does not, and should not,
change the character of the assumed underlying theory, albeit a
strongly interacting one, e.g., by inducing a prominent vector
contribution.

\showsectIDfalse
\section{Acknowledgements}
We are grateful to D. Chao, C.B. Chiu, D. Dicus, E.C.G. Sudarshan, S.
Thomas, and U. van Kolck for useful discussions and suggestions.
This work was supported in part by the Department of Energy 
Grant No. DEFG05-85ER40200.

\nosechead{References}
\ListReferences
\vfill\supereject                                     

\epsfxsize 3in\epsfbox[18 70 330 400]{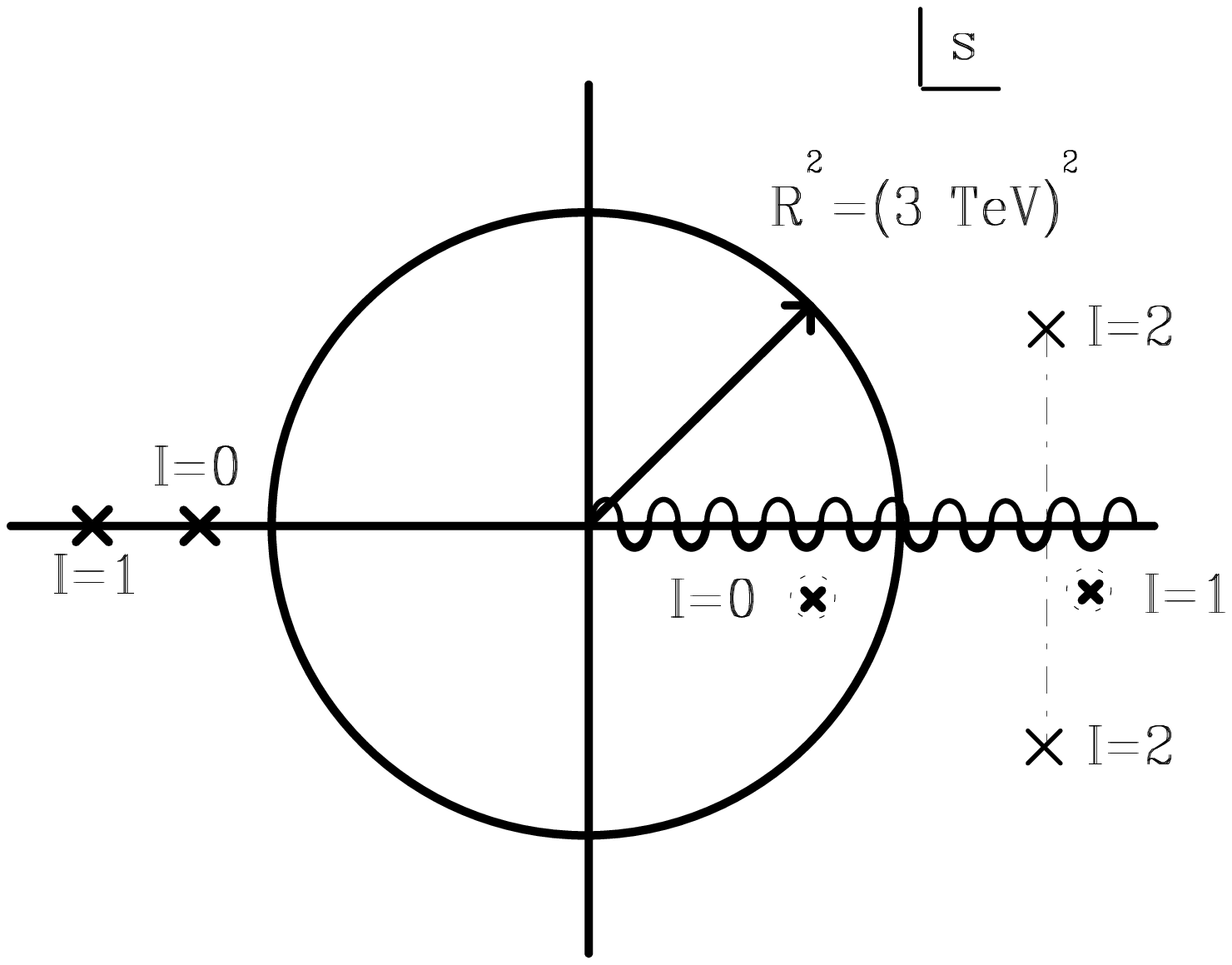}
\vskip0.25in
{\bf Figure 1:}
Schematic depiction of the complex s-plane for characteristic values of the
input parameters, M$_H$=M$_f$= 1 TeV. The only pole below the cutoff is
the physical Higgs pole.
\vskip1in
\epsfxsize 5in\epsfbox{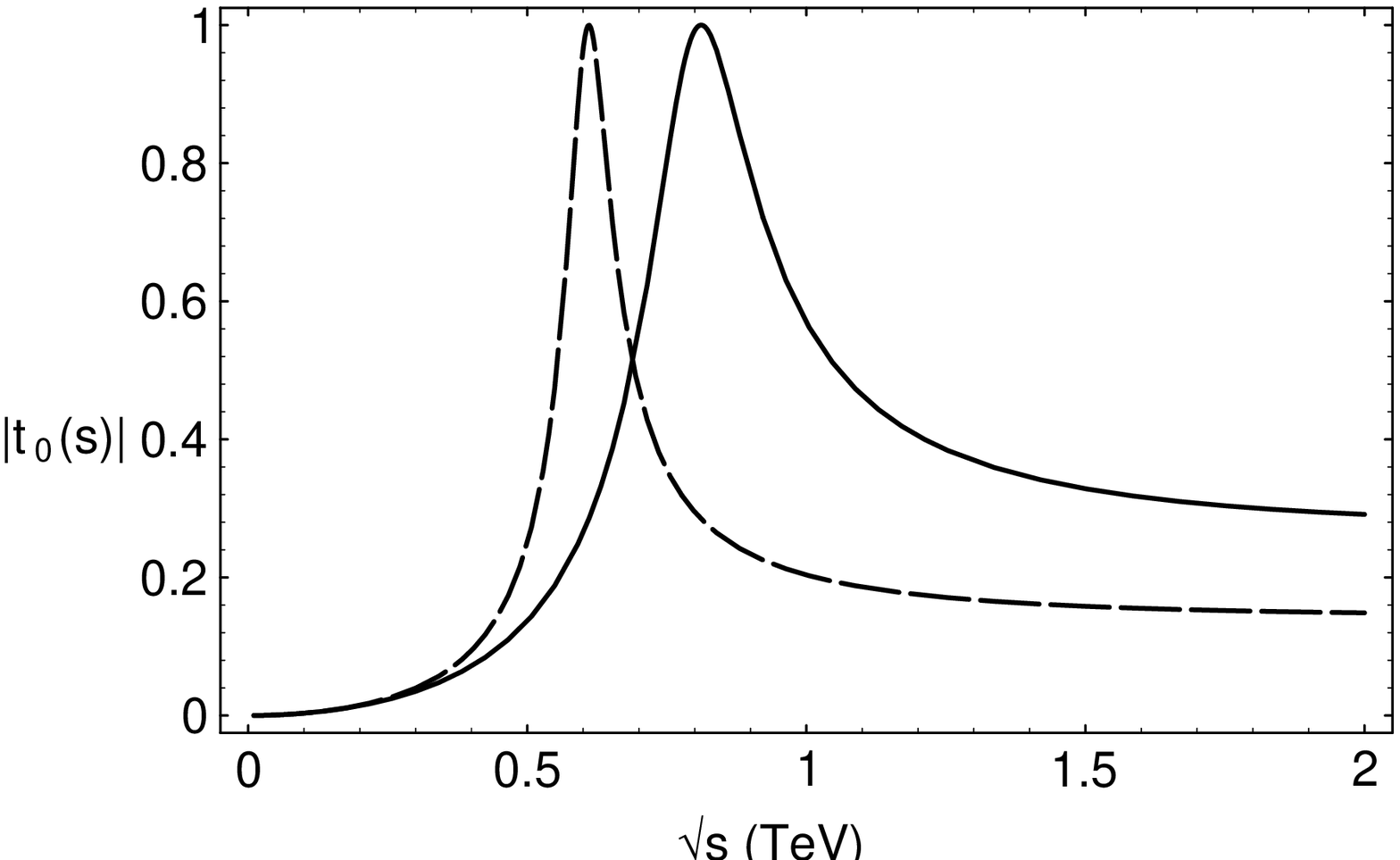}
\vskip0.25in
{\bf Figure 2a:}
I=0 s-wave amplitude.  The dashed line corresponds to 
M$_H$=0.75 TeV and the solid line to M$_H$=1 TeV.
\vfill\supereject

\epsfxsize 5in\epsfbox{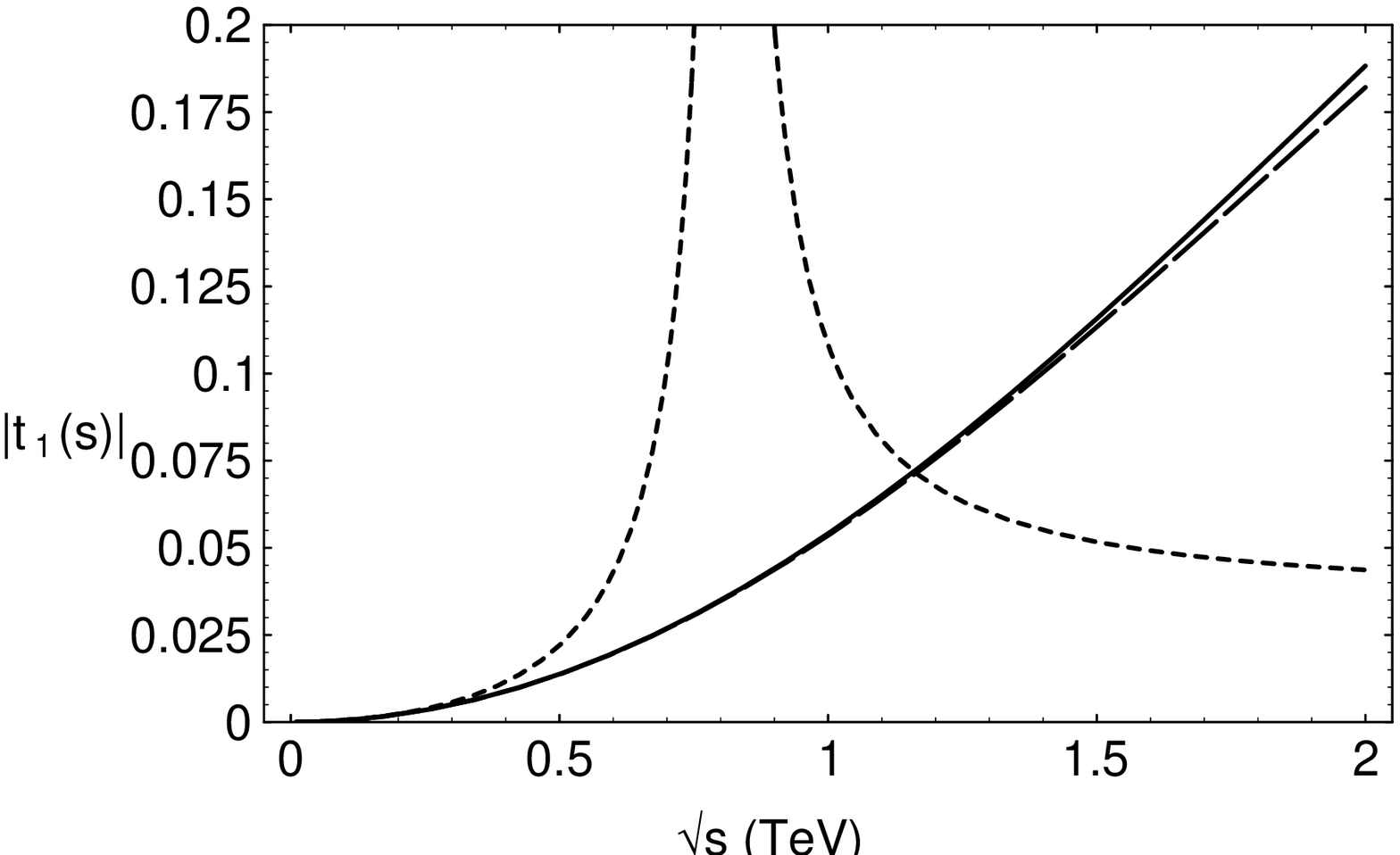}
\vskip0.25in
{\bf Figure 2b:}
I=1 p-wave amplitude.  The dashed line corresponds to 
M$_H$=0.75 TeV and the solid line to M$_H$=1 TeV. The dotted 
line corresponds to the Pad\'e method prediction for M$_H$=0.75 TeV 
and M$_f$=1 TeV
\vskip1in
\epsfxsize 5in\epsfbox{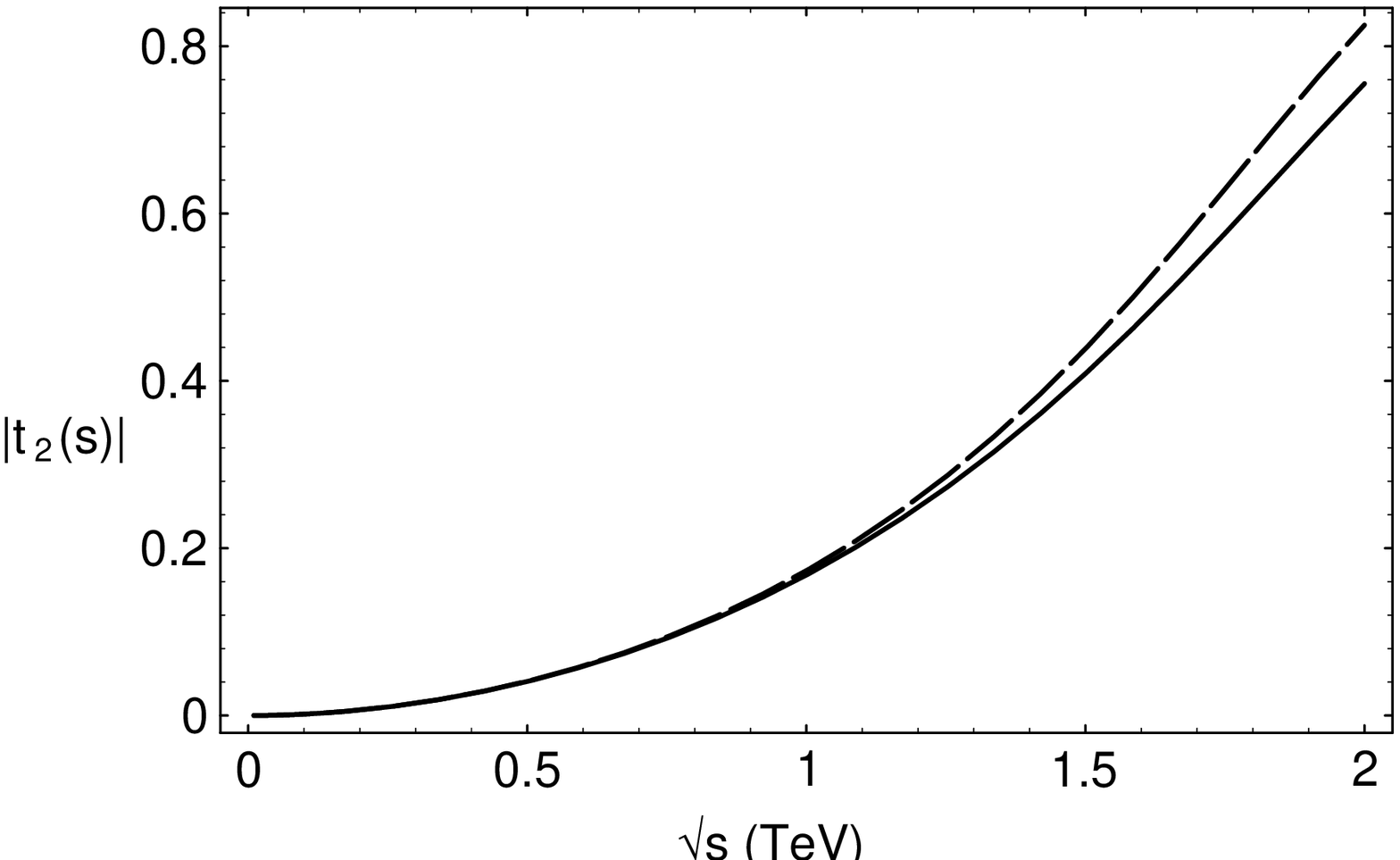}
\vskip0.25in
{\bf Figure 2c:}
I=2 s-wave amplitude.  The dashed line corresponds to 
M$_H$=0.75 TeV and the solid line to M$_H$=1 TeV.
\vfill\supereject
\end